\documentclass[%
 reprint,
superscriptaddress,
footinbib,
amsmath,amssymb,
prl,
]{revtex4-1}

\usepackage[english]{babel}
\usepackage[utf8x]{inputenc}
\usepackage[colorinlistoftodos]{todonotes}
\usepackage{enumerate}
\usepackage{graphicx}
\usepackage{dcolumn}
\usepackage{bm}

\usepackage{textcomp}
\usepackage[pagewise,columnwise]{lineno}

\begin{document}

\preprint{APS/123-QED}

\title{
Percolation leads to finite-size effects on the transition temperature and center of mass energy required for the quark-gluon plasma formation
}

\author{J. C. Texca García}
\affiliation{Facultad de Ciencias F\'isico Matem\'aticas, Benem\'erita Universidad Aut\'onoma de Puebla, Apartado Postal 165, 72000 Puebla, Pue., M\'exico}

\author{D. Rosales Herrera}
\affiliation{Facultad de Ciencias F\'isico Matem\'aticas, Benem\'erita Universidad Aut\'onoma de Puebla, Apartado Postal 165, 72000 Puebla, Pue., M\'exico}

\author{J. E. Ram\'irez}
\email{jhony.ramirezcancino@viep.com.mx}
\affiliation{Centro de Agroecología,
Instituto de Ciencias,
Benemérita Universidad Autónoma de Puebla, Apartado Postal 165, 72000 Puebla, Pue., M\'exico}
\affiliation{Consejo de Ciencia y Tecnología del Estado de Puebla, Privada B poniente de la 16 de Septiembre 4511, 72534, Puebla, Pue., México.}

\author{A. Fern\'andez T\'ellez}
\affiliation{Facultad de Ciencias F\'isico Matem\'aticas, Benem\'erita Universidad Aut\'onoma de Puebla, Apartado Postal 165, 72000 Puebla, Pue., M\'exico}

\author{C. Pajares}
\email{pajares@fpaxp1.usc.es}
\affiliation{
Departamento de F\'isica de Part\'iculas and Instituto Galego de Física de Altas Enerxías, Universidad de Santiago de Compostela, E-15782 Santiago de Compostela, Espa\~na}

\begin{abstract}
We investigate the finite-size effects on the transition temperature associated with the quark-gluon plasma (QGP) formation. From a percolation perspective, the onset of the QGP in high-energy collisions occurs when the spanning cluster of color strings emerges.
The principal result presented here is the finite-size effects on the transition temperature expressed as a power law in terms of the nucleon number.
We found that the transition temperature is higher for small systems than for large ones.
It means that minimal triggering conditions events in pp collisions require about twenty times higher energies than AuAu-PbPb collisions.
We also estimate the center of mass energy required for the QGP formation as a function of the nucleon number. Our results are consistent with the minimal center of mass energies at which the QGP has been observed.
\end{abstract}
\maketitle



In high energy physics, multiparticle production is usually described in terms of color strings stretched between the projectile and target, which decay into new strings through color neutral $q\bar{q}$ pairs production and subsequently hadronize to produce the observed hadrons \cite{BRAUN2015,string}.
Color strings may be viewed as small circular areas distributed in the transverse plane (of the collision) filled with a color field created by colliding partons that interact when they overlap.
With the increasing energy and size of the colliding system, the number of strings grows. 
They start to overlap, forming clusters in the transverse plane, in a similar way to the phenomena studied by 2-dimensional percolation theory \cite{PRLPajares}.
At one critical density, a giant cluster of color strings appears where the quarks are no more confined.  This interpretation marks the percolation phase transition and the onset of the quark-gluon plasma (QGP) \cite{chernodub}.
A first experimental evidence of the QGP was observed in AuAu collisions at RHIC that later was confirmed in PbPb collisions at LHC, through the study of all the harmonics of the azimuthal distributions, showing the existence of a collective motion of quarks and gluons \cite{intro1, intro2, intro3}. 
Furthermore, the data analysis for pp and pA collisions at LHC \cite{CMS2} and d-Au and $^3$He-Au collisions at RHIC \cite{Aidala2019} conclude that in these experiments, most properties observed in heavy-ion collisions are also present.

The general result due to the SU(3) random summation of color charges of overlapped strings is a reduction of the multiplicity and an increase of the string tension, hence an increase of the mean transverse momentum \cite{pajares2005}.
For a cluster formed by $n$ strings, it is found that its multiplicity $\mu_n$ and average transverse momentum square $\langle p_T^2 \rangle_n$ are given by \cite{Braun2000}
\begin{linenomath}
\begin{align}
    \mu_n&=(nS_n/S_1)^{1/2}\mu_1,\label{eq:mun}\\
    \langle p_T^2 \rangle_n&=(nS_1/S_n)^{1/2}\langle p_T^2 \rangle_1,\label{eq:ptn}
\end{align}
\end{linenomath}
where $S_1=\pi r_0^2$ is the transverse area of the strings, and $S_n$ the area covered by the cluster, 
$\mu_1$ and $\langle p_T^2 \rangle_1 $ are the multiplicity and average transverse momentum squared for a single string, respectively.
Here $r_0$ is the radius of the string and takes values between 0.2 fm and 0.3 fm.
Taking into account fluctuations of the number of strings at a fixed density, the average of $nS_1/S_n$ is
\begin{equation}
\left \langle nS_1/S_n \right \rangle=\eta/\phi(\eta)=F(\eta)^{-2},
\label{eq:defF}
\end{equation}
with $\phi(\eta)$ being the area covered by strings and $\eta=NS_1/S$ the filling factor, where $N$ is the total number of strings distributed on the transverse plane with area $S$ \cite{BRAUN2015,string}.
In particular, for the case of uniformly distributed disks, this picture corresponds to the classical two dimensional continuum percolation of disks, and $\phi(\eta)=1-e^{-\eta}$ in the thermodynamic limit \cite{BRAUN2015,vicsek}.
Nevertheless, $\phi(\eta)$ can significantly deviate from $1-e^{-\eta}$ if the model considers a low number of strings \cite{RAMIREZ2017}, fluctuations in the shape of the transverse surface to the collisions \cite{RAMIREZ2019}, interactions between strings \cite{RAMIREZ2021}, or non-uniform string density profiles \cite{RODRIGUES1999402}.
Equation~\eqref{eq:defF} defines the function
\begin{equation}
F(\eta)=[\phi(\eta)/\eta]^{1/2},
\label{eq:F}
\end{equation}
called the color suppression factor, which reflects the effects of the interaction between strings on the observables of the system \cite{Braun2000}.
Combining Eqs.~\eqref{eq:mun}, \eqref{eq:ptn}, and \eqref{eq:F}, we get \cite{PAJARES2011125}
\begin{linenomath}
\begin{align}
    \mu&=NF(\eta)\mu_1, \label{eq:mu}\\
    \langle p_T^2 \rangle&=\langle p_T^2 \rangle_1/F(\eta) \label{eq:pT}.
\end{align}
\end{linenomath}
Note that for diluted systems, $\eta\to 0$ and $F(\eta)\to 1$, hence $\mu=N\mu_1$ and $\langle p_T^2 \rangle=\langle p_T^2 \rangle_1$.
However, since $F(\eta)$ is a decreasing function, a colliding system produces less charged particles with higher momentum as $\eta$ increases, which can be understood as one effect of the string clustering process.
Additionally, equation \eqref{eq:pT} encodes the dependence on the centrality, energy and sizes of the collision of the average of transverse momentum of produced particles. This determines the slope of the low $p_T$ distribution and thus the radial flow. A detailed study of the comparison with data of the $p_T$ distribution and fluctuations from a color string percolation approach can be seen in Refs.~\cite{ddeus,BRAUN201314}. 

On the other hand, it is well-known that the Schwinger mechanism $dN/dp_T^2\sim e^{-\pi p_T^2/x^2}$ dictates the transverse momentum distribution of the produced particles \cite{Schwinger1, Schwinger2, Schwinger3}, with $x$ being the string tension (color field), which is expected to fluctuate.
Assuming a Gaussian distribution for the chromoelectric field \cite{BIALAS1999301} with variance $\langle x^2 \rangle$, the Schwinger distribution becomes $dN/dp_T^2\sim e^{-\beta p_T}$,
where $\beta=(2\pi/\langle x^2 \rangle)^{1/2}$ \cite{DIASDEDEUS2006455}.
Note the similarity between the latter and the Boltzmann distribution. 
In this way, $\beta$ can be understood as a measure of the inverse of the temperature of the system \footnote{The temperature in Eq.~\eqref{eq:temp} is equivalent to the one of the blast wave model, which is split into two contributions. One is the freeze-out temperature that increases with multiplicity. The remaining contribution comes from a mass-dependent term representing a radial flow.}.
To determine $\langle x^2 \rangle$, it is necessary to compute $\langle p_T^2\rangle$ from the Schwinger mechanism. Then averaging over the tension fluctuations, we get $\langle p_T^2\rangle=\langle x^2 \rangle/\pi$.
Using Eq.~\eqref{eq:pT}, we can define a dimensionless temperature for the color string percolation model as
\begin{equation}
    T^*(\eta)=\langle p_T^2 \rangle_1^{-1/2}T(\eta)=[2F(\eta)]^{-1/2}.
\label{eq:temp}
\end{equation}
This local temperature becomes the temperature of the thermal distribution as clusters grow and cover most of the collision surface \cite{DIASDEDEUS2006455}.
So the quark-gluon plasma formation can be associated with the emergence of the spanning cluster \cite{SATZ20013, GATTRINGER2010179, GattringerPRD}
and then the transition temperature can be defined as the evaluation of the temperature \eqref{eq:temp} at the percolation threshold ($\eta_c$), i.e., $T_c=T(\eta_c)$.
The value of $\langle p_T^2 \rangle_1$ can be determined by comparing $T_c$ in Eq.~\eqref{eq:temp} with the estimation made from other models, for example, lattice QCD \cite{temp}.
However, the dimensionless temperature $T_c^*$ may be helpful in the analysis of the behavior of observables through the order parameter $\epsilon=(T-T_c)/T_c=(T^*-T_c^*)/T_c^*$
which is independent of $\langle p_T^2 \rangle_1$ \cite{RAMIREZ2019}.

This paper aims to analyze the finite-size effects of the transition dimensionless temperature as a function of the system size. 
To do this for finite systems, we estimate the percolation threshold and the area covered by disks by computer simulation, then we compute the transition temperature and the minimal center of mass energy required for the formation of the QGP on pp and AA collisions.
The main result presented here is that such finite-size effects are expressed as a power-law in terms of the nucleon number. 
This lets us estimate $T_c^*$ for different projectiles in the collisions, from pp to PbPb.

In our simulations, we adopt the microcanonical ensemble scheme developed by Mertens and Moore in Ref.~\cite{mertens} for continuum percolation systems, so that strings are added one by one in a square surface of side $L$.
We then measure the value of the observable $O$ after adding exactly $n$ strings, which is denoted by $O_n$.
Thus, the mean value of the observable $O$ at any value of $\eta$ is estimated by convoluting the set of values  $\{O_n\}$ and the Poisson distribution with average $\alpha=\eta L^2/\pi r_0^2$ \cite{mertens}, i.e.,
\begin{equation}
O(\eta)=e^{-\alpha}\sum_{n=0}^\infty O_n \frac{\alpha^n}{n!}. \label{eq:av}
\end{equation}
We compute Poisson weights $w_n\propto\alpha^n/n!$ by using the recursive formula provided by Mertens and Moore in Ref. \cite{mertens}.
By plugging the $w_n$ distribution instead of Poisson weights in Eq.~\eqref{eq:av}, it is necessary to normalize the sum by dividing by $\sum w_n$.

We apply this algorithm to estimate the percolation threshold, area covered by disks, color reduction function, and temperature for color string percolation systems at finite transverse plane.


To compute the percolation threshold, we randomly add disk by disk over the square $[0,L]\times[0,L]$ until the appearance of the spanning cluster.
The Union-Find algorithm is used for the clustering process.
The spanning cluster is detected when at least one string at the left border and other at the opposite side belong to the same cluster.
After the emergence of the spanning cluster, the simulation is stopped, from which we obtain the critical number $n_c$ of disks added.
We compute the probability $f_n$ that a spanning cluster appears after adding exactly $n$ disks as the quotient between the frequency of $n$ and the total of trials performed.
Thus $\mathcal{F}_n=\sum_{k=0}^nf_k$
defines the probability of observing the spanning cluster after adding $n$ disks in the system.
Therefore, we compute the percolation probability $P(\eta)$ at any filling factor by convoluting the distribution $\mathcal{F}_n$ with the Poisson distribution as in Eq.~\eqref{eq:av}.
The percolation probability has a sigmoid form whose width transition becomes smaller as $L$ increases. Moreover, in the thermodynamic limit, $P(\eta)$ is a Heaviside step function $H(\eta-\eta_c)$, with $\eta_c$ the percolation threshold \cite{SABERI20151}.
In this way, the percolation probability for finite $L$ is well-fitted to the function
$P_L(\eta)=0.5\left[ 1+\tanh \left( (\eta-\eta_{cL})/\Delta_L  \right)  \right]$.
The subscript in $P$ and $\eta_c$ point out their dependence on $L$.
$\Delta_L$ is the width of the sigmoid transition for finite size systems \cite{Rintoul_1997,SABERI20151}.
The estimation of $\eta_c$ in the thermodynamic limit  requires a finite-size effects analysis on $\eta_{cL}$ \cite{mertens, Ziff1, Ziff2}.

\begin{figure}[ht]
\centering
\includegraphics[scale=1]{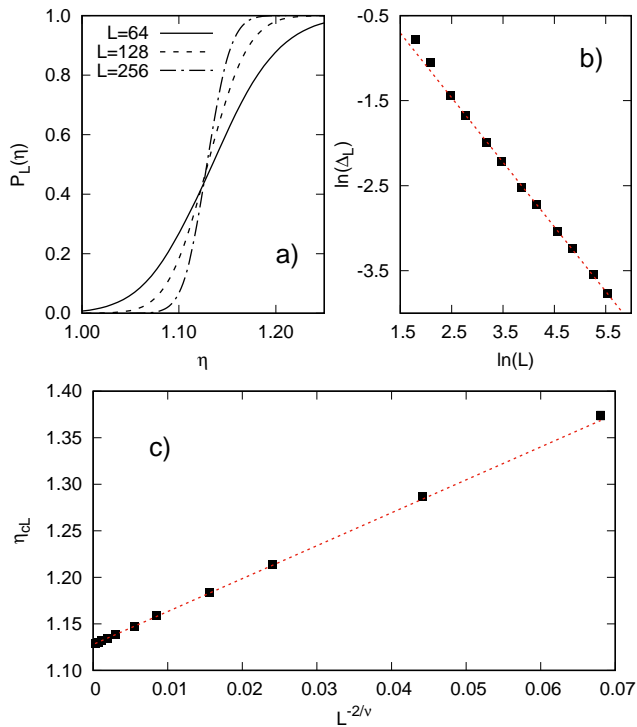}
\caption{a) Percolation probability for finite-size systems with $L=$ 64 (solid line), 128 (dashed line), and 256 (dash-dotted line). b) Width transition $\Delta_L$ (squares) of the percolation probability as a function of the system side $L$. The dashed line is the function $f(L)=aL^b$, with $b=-0.758(7)$. c) Results of $\eta_{cL}$ as a function of $L$. Note the suitable match between our results and the function $mL^{-2/\nu}+\eta_c$ (dashed line), with $\eta_c=$1.1279(1) being the estimation of the percolation threshold in the thermodynamic limit.}
\label{fig:Pperc}
\end{figure}


On the other hand, we use a grid sample method to compute the area covered by disks $\phi_n$ after adding exactly $n$-strings.
Our selection of the grid spacing is $r_0/20$.
Thus, $\phi_n$ is approximated as $\mathcal{N}_nr_0^2/400L^2$,
where $\mathcal{N}_n$ counts only those cells whose center lies on a disk after adding exactly $n$-strings.
It is expected that $\phi_n$ be a random variable because the center of the strings are randomly allocated.
We thus define $\bar{\phi}_n$ as the average of $\phi_n$ over the number of trials performed.
Therefore, the fraction of covered area by disks at any filling factor value is estimated using the convolution in Eq.~\eqref{eq:av}.
Since the distribution $\bar{\phi}_n$ is estimated for a finite number of strings, we perform the convolution for a finite number of terms.
The sum runs from $N_\text{min}=\bar{n}-5\sigma$ to $N_\text{max}=\bar{n}+5\sigma$, where $\bar{n}=\lfloor \alpha \rfloor$ and
$\sigma=\lfloor \alpha^{1/2}  \rfloor$.
Therefore, the critical area covered by disks is computed as
\begin{equation}
\phi_{cL}=e^{-\alpha_{cL}}\sum_{n=N_\text{min}}^{N_\text{max}} \bar{\phi}_n w_n \left / \sum_{n=N_\text{min}}^{N_\text{max}} w_n \right.\label{eq:avwn},
\end{equation}
with $\alpha_{cL}=\eta_{cL}L^2/\pi r_0^2$.
In what follows, we use the latter scheme to calculate the value of the color suppression factor  and dimensionless temperature at the percolation threshold just by replacing $\bar{\phi}_n$ by $F_n$ and $T_n^*$, which are computed as $(\bar{\phi}_n L^2/n\pi r_0^2)^{1/2}$ and $(2F_n)^{-1/2}$, respectively.

In the simulations, we set the diameter of the strings as the typical length of the system.
This allows us to represent the length side of the system as integer multiples of the string diameter.
To take into account the finite-size effects on the percolation threshold and other observables of interest, simulations were carried out by setting $L=$6, 8, 12, 16, 24, 32, 48, 64, 96, 128, 192, and 256.
The data analysis is achieved over a compilation of 10$^4$ trials for each $L$-value.


Figure \ref{fig:Pperc} a) illustrates the percolation probability sigmoid shape.
Notice that the width transition $\Delta_L$ becomes smaller as $L$ increases. 
This fact leads to the well-known scaling relation $\Delta_L\propto L^{-1/\nu}$, with $\nu$ the critical exponent associated to the (cluster radii) correlation length \cite{stauffer}. This behavior is depicted in Fig.~\ref{fig:Pperc} b), where $\Delta_L$ is well described by a linear function in a logarithmic scale whose slope corresponds to the critical exponent.
We found $1/\nu=0.758(7)$ for the systems described above. 
This value agrees with the existing result for percolation systems in two-dimensions, where $1/\nu=0.75$ \cite{stauffer}.

The percolation threshold in the thermodynamic limit is determined by analyzing the scaling relation of $\eta_c-\eta_{cL}$
 as a function of $L$, which is determined by two factors.
The first comes from the fact that the width transition scales as $L^{-1/\nu}$.
The second factor comes from the imposed boundary conditions, which contributes another $L^{-1/\nu}$.
These two factors combine to give
\begin{equation}
\eta_c-\eta_{cL}\propto L^{-2/\nu},
\label{eq:ptL}
\end{equation}
leading to stronger finite-size effects than the universal scaling relation for the percolation threshold.
Figure~\ref{fig:Pperc} c) shows the behavior of $\eta_{cL}$ as a function of $L^{-2/\nu}$.
Notice that our results on the percolation threshold for finite systems satisfy the scaling relation \eqref{eq:ptL} even for small values of $L$.
The percolation threshold in the thermodynamic limit ($L\to\infty$) is determined by extrapolating the trendline of the data.
We found $\eta_c=1.1279(1)$, which is in accordance with the most precise value reported in the literature (1.12808737(6)) \cite{mertens}.

\begin{figure}[ht]
\centering
\includegraphics[scale=1]{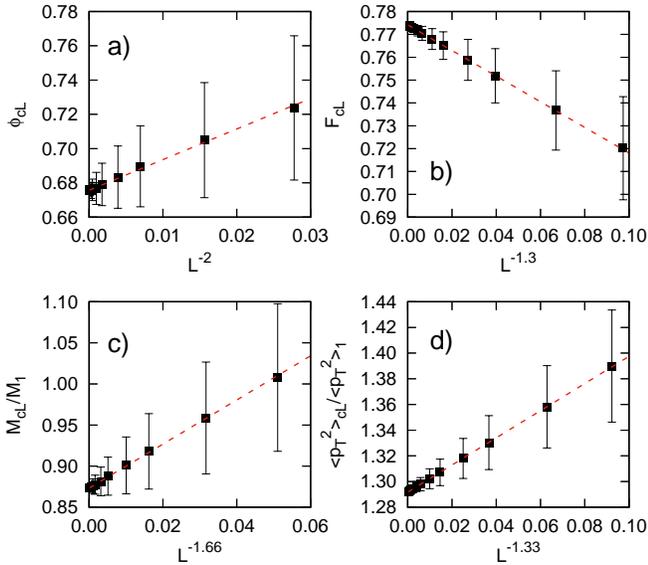}
\caption{
Area covered by disks [a)], color suppression factor [b)], multiplicity density [c)], and average transverse momentum squared [d)] at $\eta_{cL}$ as a function of $L$.
Dashed lines are the corresponding scaling relation $mL^a+b$, with $b=$0.6757(7), 0.7742(1), 0.8731(6), and 1.2917(2) being the estimation in the thermodynamic limit for $\phi_c$, $F_c$, $M_c/M_1$, and $\langle p_T^2 \rangle_c/\langle p_T^2 \rangle_1$, respectively. Error bars are computed as the square root of the variance.}
\label{fig:CAD}
\end{figure}

In addition, from the analysis of the finite-size effects, we found the following power-law behaviors for the area covered by disks and critical color suppression factor
\begin{linenomath}
\begin{align}
    \phi_c-\phi_{cL}&\propto L^{-2}, \label{eq:sc_CAD}\\
    F_c-F_{cL} & \propto L^{-1.3}, \label{eq:sc_F}
\end{align}
\end{linenomath}
respectively.
Moreover, the average transverse momentum squared at the percolation threshold behaves as
\begin{equation}
    \langle p_T^2 \rangle_c - \langle p_T^2 \rangle_{cL} \propto L^{-1.33}. \label{eq:sc_pt}
\end{equation}
Note that the multiplicity diverges as $L^2$ because $\mu$ is proportional to the number of strings in the system. By replacing $N=\eta L^2/S_1$ in Eq.~\ref{eq:mu}, we can define the multiplicity density as follows $M=\mu /L^2 = \eta F(\eta) M_1$, with $M_1=\mu_1/S_1$ being the multiplicity density for a single string.
Thus, we found
\begin{equation}
    M_c - M_{cL} \propto L^{-1.66}. \label{eq:sc_mu}
\end{equation}
Figures \ref{fig:CAD} a), b), c) and d) show the scaling relations in Eqs. \eqref{eq:sc_CAD}, \eqref{eq:sc_F}, \eqref{eq:sc_pt}, and \eqref{eq:sc_mu}, respectively.
Error bars are computed as the usual standard deviation, where the second moment 
is calculated according to the convolution in Eq.~\eqref{eq:avwn}.

Our estimation of the critical area covered by disks in the thermodynamic limit is $\phi_c=$0.6757(7), which agrees with the best existing determination (0.67634831(2)) \cite{mertens}.
Moreover, we estimate $F_c=$0.7742(1), 
$M_c/M_1=0.8731(6)$, and $\langle p_T^2 \rangle_c/\langle p_T^2 \rangle_1=1.2917(2)$ in the thermodynamic limit.
These values are in agreement with those computed using the information of $\eta_c$ and $\phi_c$ reported in Ref.~\cite{mertens}, whose estimations are  $F_c=$0.77430816(4), $M_c/M_1=$0.87348726(6), and $\langle p_T^2 \rangle_c/\langle p_T^2 \rangle_1=$1.29147548(7).

\begin{figure}[ht]
\centering
\includegraphics[scale=1]{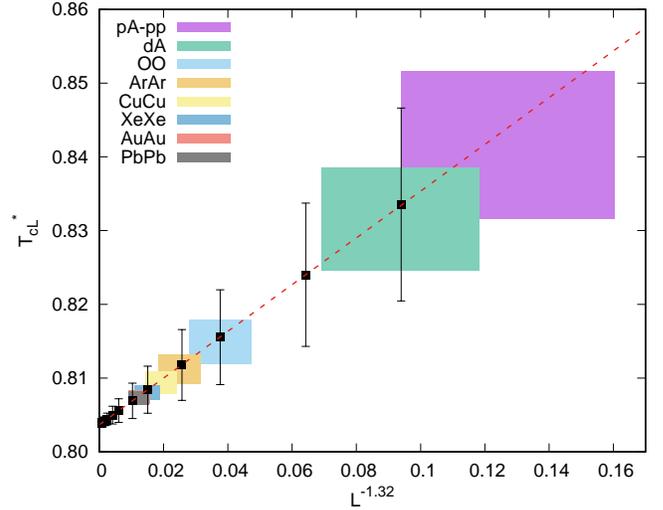}
\caption{Transition dimensionless temperature (black squares) for the estimated percolation threshold $\eta_{cL}$. Error bars are calculated as the square root of the variance. Dashed line is the function $mL^{-1.32}+T_c^*$, with $T_c^*$=0.80365(8) being the estimation of the dimensionless temperature in the thermodynamic limit. Color boxes are estimations and error propagation of the transition temperature of the quark-gluon plasma formation for pp, pA and AA collisions.}
\label{fig:Tc}
\end{figure}

We recall that the transition temperature associated with the quark-gluon plasma formation can be computed by evaluating $T$ at the percolation threshold.
For the color string percolation model we found that $T_{cL}^*$ satisfy
\begin{equation}
    T_c^*-T_{cL}^*\propto L^{-1.32},
\label{eq:fsTc}
\end{equation}
where $T_{cL}^*$ is computed by means of the convolution in Eq.~\eqref{eq:avwn}.
Figure~\ref{fig:Tc} depicts this scaling behavior for our data simulation of $T_{cL}^*$.
Another way to calculate $T_{cL}^*$ consists of computing the percolation probability as a function of the temperature.
Since $P_L(T)$ also exhibits a sigmoid shape, $T_{cL}^*$ is estimated by using the aforementioned procedure.
Both methods give consistent results on $T_{cL}^*$, which coincide up to five decimals.
Additionally, we estimate $T_c^*=$0.80365(8) in the thermodynamic limit.
As expected, our estimation perfectly agrees with the value 0.80357808(3) computed using the best determination of the percolation threshold and the critical area covered by disks reported in Ref. \cite{mertens}.
Notice the scaling relation in \eqref{eq:fsTc} implicates that the transition temperature of the quark-gluon plasma formation is greater for small systems than for large ones.

To make contact with the experimental parameters of collision physics, we enclose the transverse plane surface for central collision with a square of side length $L=R_A/r_0$, with $R_A$ being the radius of the atomic nucleus, given by $R_A=r_0^*A_M^{1/3}$,
where $A_M$ is the nucleon number and $r_0^*$ takes values between 1.2-1.3 fm \cite{krane1991}. 
Therefore, the finite-size effects on the transition temperature can be rewritten as a power law on the nucleon number as  
\begin{equation}
T_c^*-T_{cL}^*\propto A_M^{-0.44}.
\end{equation}
This last relation is helpful because we can estimate the transition temperature for different kinds of projectiles, as Fig.~\ref{fig:Tc} illustrates.
In the computation, we use $r_0^*=1.25(5)$ fm and $r_0=0.25(5)$ fm.
The error propagation leads to $\sigma_L\approx 0.2L$ and $\sigma_{T_{cL}}\approx 0.08L^{-1.32}$, which are shown as box errors in Fig.~\ref{fig:Tc}.
We take $A_M=1$ and $A_M=2$ for pA-pp and dA collisions, respectively.
For AA collisions, we take for $A_M$ the value that corresponds to the most abundant isotope.

\begin{figure}[ht]
\centering
\includegraphics[scale=1]{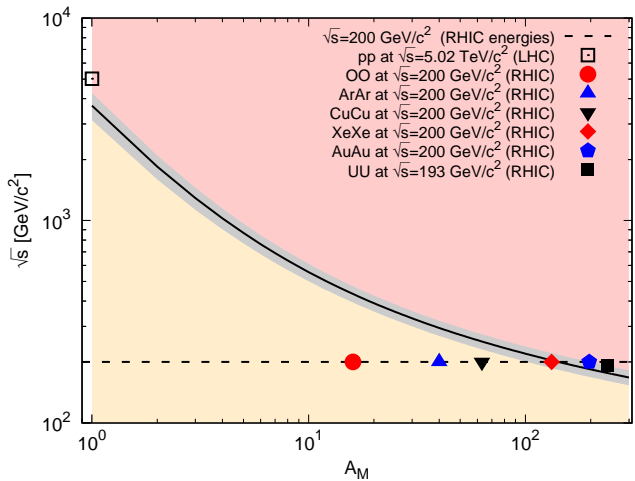}
\caption{Diagram of the QGP formation as a function of the variables $\sqrt{s}$ and $A_M$. The critical curve (solid line) marks the minimum center of energy required to observe the QGP for a collision involving $A_M$ nucleons per projectile.
The errors (grey shaded region) are computed by assuming a normal distribution for all the parameters but keeping fixed the ratio $r_0/r_0^*=0.2$. Red shaded region corresponds to the conditions of $\sqrt{s}$ and $A_M$ at which the QGP may be observed. Figures are AA and pp collision experiments at different center of mass energies.}
\label{fig:cme}
\end{figure}

Additionally, it is possible to determine the center of mass energy needed for the formation of the QGP. To do this, we consider that the filling factor of strings distributed on the transverse plane of pp collisions at center of mass energy $\sqrt{s}$ is given by \cite{BAUTISTA2012}
\begin{equation}
\eta^{pp}=\frac{\pi}{25}\left[ 2+4\left( \frac{r_0}{R_p}   \right)^2 \left( \frac{\sqrt{s}}{m_p}  \right)^{2\lambda}  \right],
\label{eq:lambda}
\end{equation}
where $R_p$=0.84-0.87 fm and $m_p$=938 MeV/c$^2$ are the radius and mass of the proton, respectively. On the other hand, the filling factor for central AA collisions is
\begin{equation}
\eta^{AA}=\eta^{pp}A_M^{\alpha(\sqrt{s})},
\end{equation}
with
\begin{equation}
\alpha(\sqrt{s})=\frac{1}{3}\left[ 1-\frac{1}{1+\ln(\sqrt{s/s_0}+1)} \right].
\label{eq:alpha}
\end{equation}
The parameters $\lambda$ and $\sqrt{s_0}$ can be found by fitting the adequate experimental data, taking the values 245(29) GeV/c$^2$ and 0.201(3) \cite{BAUTISTA2012}, respectively. Since the onset of the QGP is associated to the emergence of the spanning cluster, the center of mass energy $\sqrt{s_c}$ for central AA collisions is estimated by solving the equation $\eta_c+mA_M^{-2/3\nu}=\eta^{AA}$, with $m=0.316(4)$.
In Fig.~\ref{fig:cme} we plot $\sqrt{s_c}$ as a function of the nucleon number. 
We estimate $\sqrt{s_c}$=184(15) GeV/c$^2$ for AuAu collisions.
Notice that our estimation is in agreement with the data analysis that claims the QGP signatures for  AuAu ($\sqrt{s}=$200 GeV/c$^2$ at RHIC) collisions \cite{GYULASSY200530}.
For PbPb collisions we found $\sqrt{s_c}$=182(15) GeV/c$^2$. This explains why the QGP is obtained at LHC energies for these experiments.
In the case of pp collisions we estimate $\sqrt{s_c}$=3.7(5) TeV/c$^2$, which is in agreement with the results observed in experiments at $\sqrt{s}=$5.02 TeV/c$^2$ at LHC \cite{2017193, atlas}. However, the formation of the QGP in pp collisions is expected for high multiplicity events at lower energies \cite{highmultiplicitypp}.


In summary, we presented the analysis of the finite-size effects on the transition temperature of the QGP formation associated with the emergence of the spanning cluster of color strings. Through the radius of the atomic nucleus, these effects are expressed as a power-law in terms of the nucleon number, which allowed us to estimate the transition temperature for different projectiles and targets in high energy collisions systems. We found that the transition temperature is higher for small collision systems like pp or pA collisions than for large ones, such as AuAu or PbPb collisions. This means that systems of color strings formed by pp or pA collisions need high filling factor values, which require higher energy or very high multiplicity than those at minimal triggering conditions events 
In particular, we estimate that pp collisions require about twenty times bigger center mass energy than AuAu or PbPb collisions for the QGP formation.

Our results could lead to finding scaling functions that incorporate information on finite-size systems and the filling factor for observables in the color string percolation framework and other models. Moreover, results presented here can be extended to consider other centralities, fluctuations of the initial shape of the transverse plane, non-uniform density profiles or interactions between strings.

\begin{acknowledgments}
J.E.R. acknowledges financial support from Consejo de Ciencia y Tecnología del Estado de Puebla.
C.P. thanks the grant Maria de Maeztu unit of excelence MDM-2016 0682 of Spain, the support of Xunta de Galicia and project PID 2020-119632GB-I00.
We also thank Nestor Armesto, Miguel Ángel García-Ariza and Bogar Díaz for their valuable comments.
\end{acknowledgments}

\bibliography{bib}

\end{document}